\documentclass[11pt]{article}
\usepackage{hyperref}
\pdfoutput=1

\begin{document}

\title{Lengths matters, periodically.\\ (The movie) \\ Fluid
Dynamics Video}
\author{M. C. Renoult, S. Ferjani, C. Rosenblatt, P. Carles \\
\\\vspace{6pt} Fast Lab., UPMC/Paris 6, FRANCE \\ Physics Dept., CWRU, Cleveland, OH,USA} \maketitle

The video shows the development of the Rayleigh-Taylor Instability
between two immiscible fluids for four distinct initial single-mode
perturbations.  To obtain such controlled initial conditions, a
Magnetic Levitation technique is used (see \cite {Ref1}). The
principle is explained in the first part of the animation. A
homogeneous magnetic force is produced in opposition to gravity and
allows the stabilization of the dense fluid (paramagnetic aqueous
mixture) above the less dense fluid (hexadecane). In addition,
segments of magnetically permeable wires are placed on the outside
of the cell in a precise configuration - predicted numerically - to
achieve almost any desired initial condition (see \cite {Ref2}). The
initial interface thus is no longer flat and the experiment is
started by turning off the magnetic field and allowing the denser
fluid to fall under gravity. Here the wires are periodically aligned
along a row, a few mm above the initial interface position,
producing a small amplitude single mode perturbation of the same
wavelength as the wires (but too small to image).  The video
presents  the Rayleigh-Taylor Instability for four decreasing
wavelengths (15 mm, 12.5, 10 mm and 7.5 mm).  In each case, one
observes the different stages of development of the Rayleigh-Taylor
Instability from the early time linear behavior to the late mixing
flow. The second instability also can be observed. \newline \newline
\noindent The link for the video is:
\href{http://ecommons.library.cornell.edu/bitstream/1813/8237/2/LIFTED_H2_EMS
T_FUEL.mpg}{Video}.

\end{document}